\documentclass[12pt]{article}
\usepackage{graphicx}
\usepackage{amsmath}
\usepackage{amssymb}
\usepackage{sidecap}
\usepackage{wrapfig}
\newcommand\pubnumber{DESY 14-217\\SFB/CPP-14-90}
\newcommand\pubdate{\today}

\def\support{\footnote{Speaker}}

\def\Title#1{\begin{center} {\Large #1 } \end{center}}
\def\Author#1{\begin{center}{ \sc #1} \end{center}}
\def\Address#1{\begin{center}{ \it #1} \end{center}}

\newcommand\pubblock{\rightline{\begin{tabular}{r} \pubnumber\\
         \pubdate  \end{tabular}}}
\newenvironment{Abstract}{\begin{quotation}  }{\end{quotation}}
\newenvironment{Presented}{\begin{quotation} \begin{center} 
             PRESENTED AT\end{center}\bigskip 
      \begin{center}\begin{large}}{\end{large}\end{center} \end{quotation}}
\def\Acknowledgements{\bigskip  \bigskip \begin{center} \begin{large}
             \bf ACKNOWLEDGEMENTS \end{large}\end{center}}
\def\beq{\begin{equation}}
\def\eeq#1{\label{#1}\end{equation}}
\def\eeqn{\end{equation}}

\def\beqa{\begin{eqnarray}}
\def\eeqa#1{\label{#1}\end{eqnarray}}
\def\eeqan{\end{eqnarray}}

\let\bar=\overbar

\def\Dslash{\not{\hbox{\kern-4pt $D$}}}
\def\dslash{\not{\hbox{\kern-2pt $\del$}}}

\def\msb{{\bar{\ssstyle M \kern -1pt S}}}

\providecommand{\mathrm}[1]{\\mathrm{#1}}
\providecommand{\V}{{\mathrm V}}
\begin{document}
\begin{titlepage}
\pubblock

\vfill
\Title{Form factors for $\mathrm B_\mathrm s \to \mathrm K \ell \nu$ decays in Lattice QCD}
\vfill
\Author{Felix Bahr\support, Fabio Bernardoni, John Bulava, Anosh Joseph,\\
Alberto Ramos, Hubert Simma, Rainer Sommer}
\Address{John von Neumann Institute for Computing (NIC), DESY, Platanenallee 6, D-15738 Zeuthen, Germany}
\vfill
\begin{Abstract}
We present the current status of the computation 
of the form factor $f_+ (q^2)$ for the semi-leptonic decay 
$\mathrm B_\mathrm s \to \mathrm K \ell \nu$
by the ALPHA collaboration.
We use gauge configurations which were generated as part of the Coordinated Lattice Simulations (CLS) 
effort. They have $N_\mathrm f=2$ non-perturbatively $O(a)$ improved Wilson fermions, and
pion masses down to $\approx 250 \,\mathrm {MeV}$ with $m_\pi L \geq 4$.
The heavy quark is treated in non-perturbative Heavy Quark Effective Theory (HQET). 

We discuss how to extract the form factors from the correlation functions 
and present first results for the form factor at $q^2=21.23\,\mathrm{GeV}^2$ 
extrapolated to the continuum. Next-to-leading order terms in HQET and 
the chiral extrapolation still need to be included in the analysis.
\end{Abstract}
\vfill
\begin{Presented}
8th International Workshop on the CKM Unitarity Triangle (CKM 2014), Vienna, Austria, September 8-12, 2014
\end{Presented}
\vfill
\end{titlepage}
\def\thefootnote{\fnsymbol{footnote}}
\setcounter{footnote}{0}

\section{Introduction}
Determinations of the CKM matrix element $|V_\mathrm{ub}|$ from different exclusive (and inclusive) 
decays tend to disagree at the $\sim 2-3\sigma$ level \cite{Agashe:2014kda}.
% The recent PDG report \cite{Agashe:2014kda} lists these as: $|V_\mathrm{ub}|= (4.41 \pm0.15^{+0.15}_{-0.19})\times 10^{-3}$ from inclusive $\mathrm B\rightarrow X_\mathrm u \ell \nu$ decays; $|V_\mathrm{ub}|= (3.28 \pm 0.29)\times 10^{-3}$ from exclusive $\mathrm B \rightarrow \pi \ell \nu$ decays; and $|V_\mathrm{ub}|= (4.22 \pm 0.42)\times 10^{-3}$ from $\mathrm B \rightarrow \tau \nu$ decays. 
Both theoretical and experimental improvements are needed to clarify the situation.

In this work, we report on our ongoing effort to non-perturbatively determine
the form factors for $\mathrm B_\mathrm s \to \mathrm K \ell \nu$ decays
from Lattice Quantum Chromodynamics (LQCD) with $N_\mathrm f = 2$ sea quarks. 
Although no experimental data
is available yet for this decay, the heavier spectator $s$-quark renders the
LQCD computations technically simpler than for $\mathrm B \to \mathrm \pi \ell \nu$,
and thus provides a good starting point to gain solid control on all systematic 
errors (and to make an LQCD prediction).

The decay amplitude for $\mathrm B_\mathrm s \to \mathrm K \ell \nu$ is proportional to
$|V_\mathrm{ub}|$ times the hadronic matrix element 
$\big\langle \mathrm K (p_\mathrm K) \big| \V_\mu \big| \mathrm B_\mathrm s(p_{\mathrm B_\mathrm s} ) \big\rangle$
of the vector current $\V_\mu (x) = \overline \psi _\mathrm u (x) \gamma _\mu \psi _ \mathrm b (x)$.
%%% \begin{equation}
%%% \big\langle \mathrm K (p_\mathrm K) \big| V^\mu \big| \mathrm B_\mathrm s(p_{\mathrm B_\mathrm s} ) \big\rangle 
%%% = {f_+(q^2)} \bigg[ p_{\mathrm B_\mathrm s}^\mu + p_\mathrm K^\mu - \frac{m_{\mathrm B_\mathrm s}^2 -m_\mathrm K^2}{q^2} q^\mu \bigg] 
%%% + {f_0(q^2)} \frac{ m_{\mathrm B_\mathrm s} ^2 - m_\mathrm K^2}{q^2} q^\mu ,
%%% \end{equation}
The matrix element is parametrised by two form factors, $f_0(q^2)$ and $f_+(q^2)$, which depend
on $q^2 = (p_{\mathrm B_\mathrm s} - p_\mathrm K)^2$, the invariant mass of the lepton pair.
In the limit of vanishing lepton masses only $f_+(q^2)$ contributes to the decay rate.

$|V_\mathrm{ub}|$ can then be determined by combining the differential decay rate from 
experiment with $f_+(q^2)$ from theory. In principle, it is sufficient to do
this at a single value of $q^2$.
In practice, experimental data is provided over a range (of bins) of $q^2$, and one 
can use the BCL paramterisation \cite{Bourrely:2008za} to express the form factor
$f_+(q^2)$ as a continuous function of $q^2$. Then, a theoretical prediction of
$f_+(q^2)$, e.g. from LQCD, for at least a single $q^2$ allows to extract 
$|V_\mathrm{ub}|$.

Here we report on preliminary work to study the feasibility of a precise determination 
of the form factor in the continuum limit and at a fixed $q^2$. 

\section{HQET on the lattice}
%%%%%%%%%%%%%%%%%%%%%%%%%%%%%
On the lattice (with spatial extent $L$ and lattice spacing $a$) the large mass of the
$b$ quark gives rise to a hierarchy of scales
\begin{equation}
  L^{-1}  \ll m_\pi \approx 140\,\mathrm{MeV} \ll m_\mathrm B \approx 5\,\mathrm{GeV} \ll a^{-1} ,
\end{equation}
which cannot be directly simulated with present computing resources. Instead, we follow the
strategy devised by the ALPHA collaboration \cite{Heitger:2003nj}
to treat the heavy quark within the framework of non-perturbative Lattice HQET.
It is an expansion in inverse powers of the heavy quark mass $m_\mathrm h$
and valid for kaon momenta $p_\mathrm K \ll m_\mathrm h$. In practice, we require
$p_\mathrm K \lesssim 1\, \mathrm{GeV}$. 
A key feature of Lattice HQET is that it is (believed to be) non-perturbatively renormalisable 
order by order in $1/m_\mathrm h$, and thus the computations have a well-defined 
continuum limit. 
The expectation value of a product of local fields, $\mathcal O$, up to and including $O(1/m_\mathrm h)$ 
in HQET on the lattice is 
\begin{equation}
 \left\langle \mathcal O \right\rangle = 
 \left\langle \mathcal O \right\rangle _\mathrm{stat} 
 + {\omega_\mathrm{kin}} a^4 \sum\limits_x \left\langle \mathcal O \mathcal O _\mathrm{kin} (x) \right\rangle _\mathrm{stat} 
 + {\omega_\mathrm{spin}} a^4 \sum\limits_x \left\langle \mathcal O \mathcal O_\mathrm{spin} (x) \right\rangle _\mathrm{stat}
 , \vspace{-0.2cm} %\\
%\omega_{\{\mathrm{kin,spin}\} } } \sim 1/m_\mathrm h
\end{equation}
where $\langle\ldots\rangle_\mathrm{stat}$ is the expectation value in the static approximation.
On the right hand side, also $\mathcal O$ needs to be expanded in $1/m_\mathrm h$, for instance, 
\begin{equation}
 \V_{k}^\mathrm{HQET} (x) = Z_{V_{k}}^\mathrm{HQET} \bigg [ \V_{k}^\mathrm{stat} (x) 
  + \sum\limits_{j=1}^{4} {c_{\mathrm V_{k,j}}} \V_{k,\,j} (x) \bigg] , \vspace{-0.2cm} 
\end{equation}
for the spatial components of the vector current, and analogous for $\V_0$.
The HQET parameters $\omega_{\mathrm kin}$, $\omega_{\mathrm spin}$, and $c_{\mathrm V_{\mu,j}}$ 
are of order $1/m_\mathrm h$, while $Z_{V_0}^\mathrm{HQET}$ and $Z_{V_k}^\mathrm{HQET}$ are of 
order $1$. They can be determined 
fully non-perturbatively by matching HQET and QCD \cite{Blossier:2012qu}. Thus, 
perturbation theory can be avoided at any stage of the computation. 

Since the non-perturbative matching is still in progress, we present in this exploratory
work only results in the static approximation, i.e. setting 
$\omega_\mathrm{kin} = \omega_\mathrm{spin} = c_{\mathrm V} = 0$.
%%% The components of the vector current in QCD are then given by \cite{Sommer:2010ic}
%%% \begin{align}
%%%  V_0^\mathrm{QCD} &= C_\mathrm{PS} (M_\mathrm h / \Lambda_{\overline{\mathrm{MS}}}) 
%%%  Z_\mathrm{A,RGI}^\mathrm{stat} (g_0) 
%%%  Z_\mathrm{V/A}^\mathrm{stat} (g_0) V_0^\mathrm{stat}, \label{v0renorm} \\
%%%  V_k^\mathrm{QCD} &= C_\mathrm{V}  (M_\mathrm h / \Lambda_{\overline{\mathrm{MS}}}) 
%%%  Z_\mathrm{A,RGI}^\mathrm{stat} (g_0) V_k^\mathrm{stat}, \label{vkrenorm}
%%% \end{align}
For the renormalisation constants we follow the lines of \cite{Sommer:2010ic,Heitger:2004gb}
and write $Z^\mathrm{HQET}$ as a product of matching factors, $C_\mathrm{PS}$ or $C_\mathrm{V}$, 
which are known at three loops in perturbation theory \cite{Chetyrkin:2003vi},
and $Z_\mathrm{A,RGI}^\mathrm{stat}$ which is known non-perturbatively \cite{Della Morte:2006sv}.
Truncation of the lattice theory at static order is expected to be a 10-20\% effect.

To perform the continuum extrapolation of the form factors at a fixed value of $q^2$, we 
employ flavour twisted boundary conditions \cite{Bedaque:2004kc} for the $s$ quark, 
$\psi(x+L\hat k) = \mathrm e ^{\mathrm i \theta_k} \psi(x)$. In this way, the
quark momentum is altered from $\vec p =  2\pi \vec n/L$ to $\vec p^\theta = (2\pi \vec n + \vec\theta)/L$, 
with $\vec n \in \mathbb N ^3$. Choosing the twist angle $\theta_k$, one can 
freely tune the momentum of the $s$ quark, and thus of the kaon. 
The heavy quark is twisted by the same angle to remain in the rest frame of the 
$\mathrm B_\mathrm s$ meson.

Our computations are performed on gauge field ensembles generated with $N_\mathrm f =2$
dynamical sea quarks within the CLS effort. They use non-perturbatively $O(a)$-improved 
Wilson fermions and the scale is set via $f_\mathrm K$ \cite{Fritzsch:2012wq}. 
All ensembles have $m_\pi L \gtrsim 4$. In this work, we 
present results from measurements on the three CLS ensembles A5, F6 and N6. 
Their properties are listed in ref. \cite{Fritzsch:2012wq}.
Error estimates take into account correlations and autocorrelations \cite{Schaefer:2010hu}. 
%Table \ref{tab:ens} lists the ensembles used in this work.
% \begin{table}
% \begin{center}
% \begin{tabular}{ccccccc}
% id & $T\times L^3$ & $a$ [fm] &  $m_\pi$ [MeV] & $m_\pi L$  & \# meas. & \# target \\ \hline
% A5  & $64 \times 32^3$  &  $0.0749(8)$         &$330$  & $4.0$ & 500 & 500 \\
% F6  & $96 \times 48^3$  &  $0.0652(6)$         &$310$  & $5.0$ & 254 & 500 \\
% N6  & $96 \times 48^3$  &  $0.0483(4)$         &$340$  & $4.0$ & 220 & 500 \\
% \end{tabular} \caption{List of ensembles used in this work.}\label{tab:ens}
% \end{center}
% \end{table}

\section{Computation of the form factor}
%%%%%%%%%%%%%%%%%%%%%%%%%%%%%%%%%%%%%%%%
On the gauge configurations we measure the two- and three-point correlation functions
\begin{gather}
 \mathcal C^{\mathrm K} (x^0-y^0; \vec p)  =  \sum\limits_{\vec x, \vec y} 
 \mathrm e^{-\mathrm i \vec p \cdot (\vec x - \vec y)} 
 \langle P^{\mathrm s \mathrm u} (x) P^{\mathrm u\mathrm s }(y)  \rangle
 , \quad
 \mathcal C^{\mathrm B}_{ij} (x^0-y^0; \vec 0)  =  \sum\limits_{\vec x, \vec y} 
 \langle P^{\mathrm s\mathrm b}_i(x) P^{\mathrm b\mathrm s}_j(y)  \rangle
 , \nonumber \\
 \mathcal C^{3}_{\mu,\,j}(t_\mathrm K, t_\mathrm B; \vec p) = \sum\limits_{\vec x_{\mathrm K}, \vec x_V, \vec x_{\mathrm B}} 
 \mathrm e^{-\mathrm i \vec p \cdot (\vec x_{\mathrm K} - \vec x _V)} 
 \langle P^{\mathrm s\mathrm u}(x_{\mathrm K}) \V_\mu(x_V) P^{\mathrm b\mathrm s}_j (x_{\mathrm B}) \rangle, 
\end{gather}
where $P^{\mathrm{q}_1 \mathrm{q}_2}_i(x)$ are interpolating fields, like 
$\overline{\psi}_{\mathrm{q}_1} (x) \gamma_5 \psi_{\mathrm{q}_2} (x)$, for the mesons. The subscripts $i$ or $j$
indicate different levels of Gaussian smearing \cite{Gusken:1989ad} of the $s$ quark in the heavy-light meson, 
i.e. different trial wave functions. In the limit of large Euclidean times, 
$t_\mathrm B \equiv x^0_{\mathrm B} - x^0_V$ and $t_\mathrm K \equiv x^0_V - x^0_{\mathrm K}$, the ratio
\begin{equation}
 f_{\mu,\,i}^\mathrm{ratio} (t_\mathrm B, t_\mathrm K; q^2) \equiv
 {\frac{\mathcal C^{3}_{\mu,\,i} (t_\mathrm K, t_\mathrm B)}
 {\sqrt{ \mathcal C ^\mathrm K (t_\mathrm K) \mathcal C^{\mathrm B}_{ii} (t_\mathrm B) }} \,  
 \mathrm e^{E_\mathrm K t_\mathrm K/2}\mathrm e^{E_\mathrm B t_\mathrm B/2}} 
\label{fratio}
\end{equation}
will then give the desired matrix element (for any suitable smearing $i$)
\begin{equation}
 \langle \mathrm K(p_\mathrm K^{\theta}) | \V_\mu | \mathrm B_\mathrm s(0 ) \rangle = 
  \lim_{T, t_\mathrm B, t_\mathrm K \to \infty}  f_{\mu,\,i}^\mathrm{ratio} (t_\mathrm B, t_\mathrm K; q^2)
 \label{fme} 
\end{equation}
Alternatively, we can parameterise the correlation functions as
\begin{gather}
 \mathcal C^{\mathrm K} (t_\mathrm K)  = \sum _m (\kappa^{(m)})^2 \mathrm e^{-E^{(m)}_\mathrm K t_\mathrm K}
 , \quad 
 \mathcal C_{{ij}}^{\mathrm B} (t_\mathrm B)  = \sum _n \beta_{i}^{(n)} \beta_{j}^{(n)} \mathrm e^{-E^{(n)}_\mathrm B t_\mathrm B}
 , \\
 \mathcal C_{\mu,\,i}^3 (t_\mathrm B, t_\mathrm K) = \sum _{m,n} \kappa^{(m)} {\varphi_\mu^{(m,n)}} \beta_{i}^{(n)} 
 \mathrm e^{-E^{(n)}_\mathrm B t_\mathrm B} \mathrm e^{-E^{(m)}_\mathrm K t_\mathrm K},
\label{fit}
\end{gather}
and determine $\{\kappa^{(m)}, E^{(m)}_\mathrm K\}$ from a fit to $\mathcal C^{\mathrm K}$,
and $\{ \beta_{i}^{(n)}, \varphi_\mu^{(n,m)}, E^{(n)}_\mathrm B\}$ from a combined
fit
%%% \footnote{We also considered one combined fit to all correlators and obtained consistent results.} 
to $\mathcal C_{\mu {i}}^3$ and $\mathcal C_{{ij}}^{\mathrm B}$.
Then, $\varphi_\mu^{(1,1)}$ is equal to the matrix element of eq.~(\ref{fme}).
We take $m=1$ and $n=1,2$, i.e. we include the first excited $\mathrm B_\mathrm s$ 
state, but only the kaon ground state.
%%% The form factor $f_\mu^\mathrm{ratio} (q^2)$ of eq. \eqref{fratio} is then given by $\varphi_\mu^{(1,1)}$.

\begin{wrapfigure}{l}{0.49\textwidth}
\begin{center}
\vspace*{0ex}
\scalebox{0.5}{\includegraphics{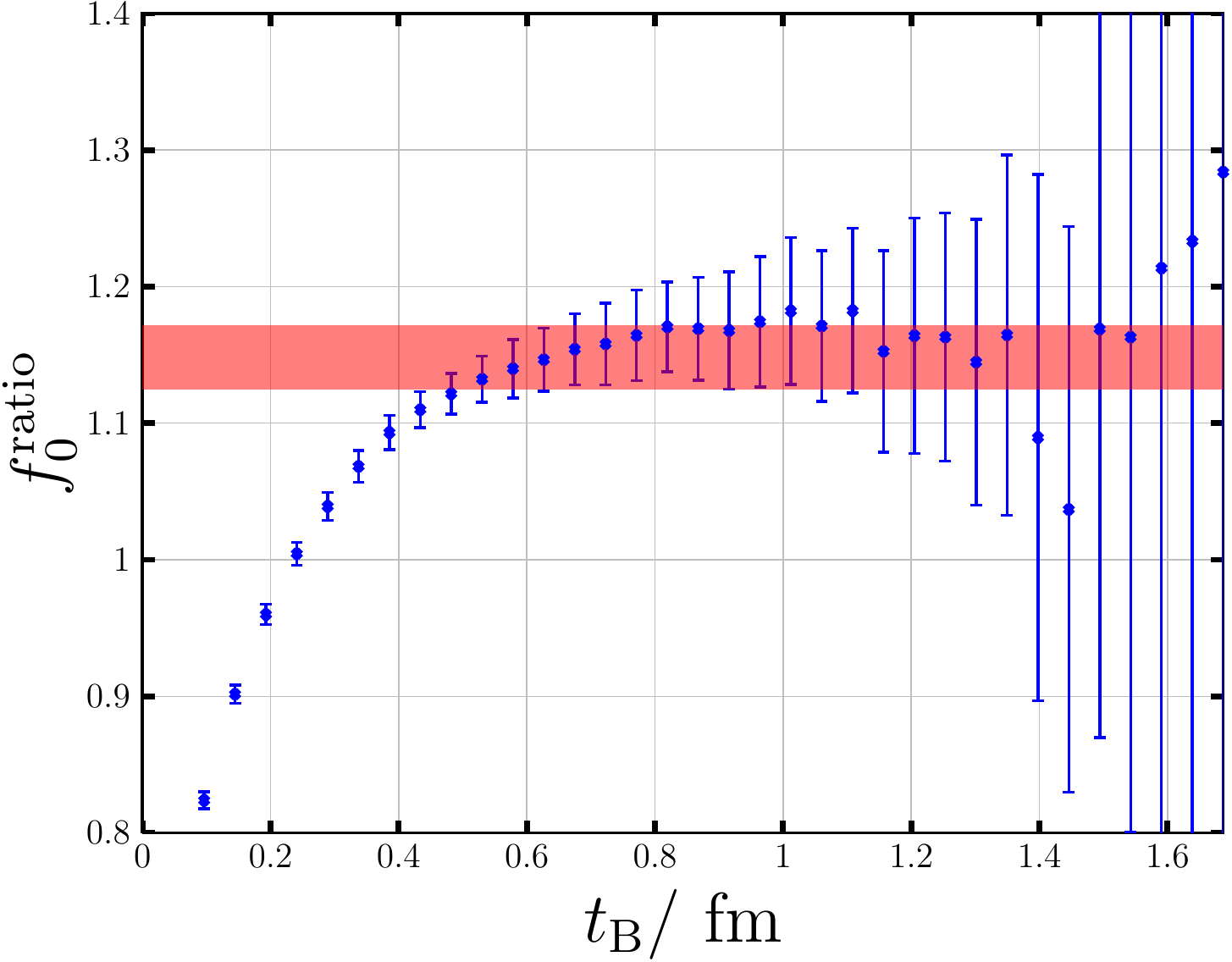}}
\caption{The ratio $f^\mathrm{ratio}_{\mu}$ (blue points) and the fit result $\varphi_\mu^{(1,1)}$ (red band) 
for lattice N6, $\mu=0$ and fixed $t_\mathrm K = 20$.
} 
\label{fig:N6}
\vspace*{-10ex}
\end{center}
%\end{figure}
\end{wrapfigure}

In fig.~\ref{fig:N6} we show the ratio $f^\mathrm{ratio}_\mu$ of eq.~\eqref{fratio} at 
fixed $t_\mathrm K=20$ as a function of $t_\mathrm B$. For comparison,
we also indicate the value of $\varphi^{(1,1)}_\mu$ resulting from the fit.

%%%%%%%%%%%%%%%%%%%%%%%%%%%%%%%%%%%%%%%%%%%%%%%%%%%%%%%%%%%%%%%%%%%%%%%%%
%%
%%   use this format to include an .eps figure into your paper
%%
% \begin{figure}[htb]
% \centering
% \includegraphics[height=1.5in]{magnet}
% \caption{Plan of the magnet used in the mesmeric studies.}
% \label{fig:magnet}
% \end{figure}
%%%%%%%%%%%%%%%%%%%%%%%%%%%%%%%%%%%%%%%%%%%%%%%%%%%%%%%%%%%%%%%%%%%%%%%%%%%
%\begin{figure}
In the rest frame of the $\mathrm B_\mathrm s$ meson, the matrix elements have the form
\begin{align}
  \langle \mathrm K | \V_0 |\mathrm B_\mathrm s\rangle 
  &= \sqrt{2 m_{\mathrm B_\mathrm s}}\, f_\parallel(q^2), \qquad
  \nonumber\\
  \langle \mathrm K | \V_i |\mathrm B_\mathrm s\rangle 
  &= \sqrt{2 m_{\mathrm B_\mathrm s}}\, p_\mathrm K ^i\, f_\perp(q^2) ,
  \nonumber
\end{align}
where the form factors $(f_\parallel, f_\perp)$ are related to $(f_+,f_0)$. 
In particular, we have
\begin{align}
  f_+ &= \frac{1}{ \sqrt{2 m_{\mathrm B_\mathrm s}}}\, f_\parallel  
     +\frac{1}{\sqrt{2m_{\mathrm B_\mathrm s}}}(m_{\mathrm B_\mathrm s}-E_\mathrm K)\, f_\perp . \label{fpl} 
  %%% +\frac{m_{\mathrm B_\mathrm s}-E_\mathrm K}{\sqrt{2m_{\mathrm B_\mathrm s}}}\,f_\perp . \label{fpl} 
  %%% \\
  %%% f_0 &= \frac{ \sqrt{2 m_{\mathrm B_\mathrm s}}}{ m_{\mathrm B_\mathrm s}^2-m_\mathrm K^2} [(m_{\mathrm B_\mathrm s}-E_\mathrm K)f_\parallel  +(E_\mathrm K^2-m_\mathrm K^2)f_\perp ] . \label{HQET_ff}
\end{align}

\begin{figure}[htbp]
\begin{minipage}[hbt]{0.49\textwidth}
		\scalebox{0.50}{\includegraphics{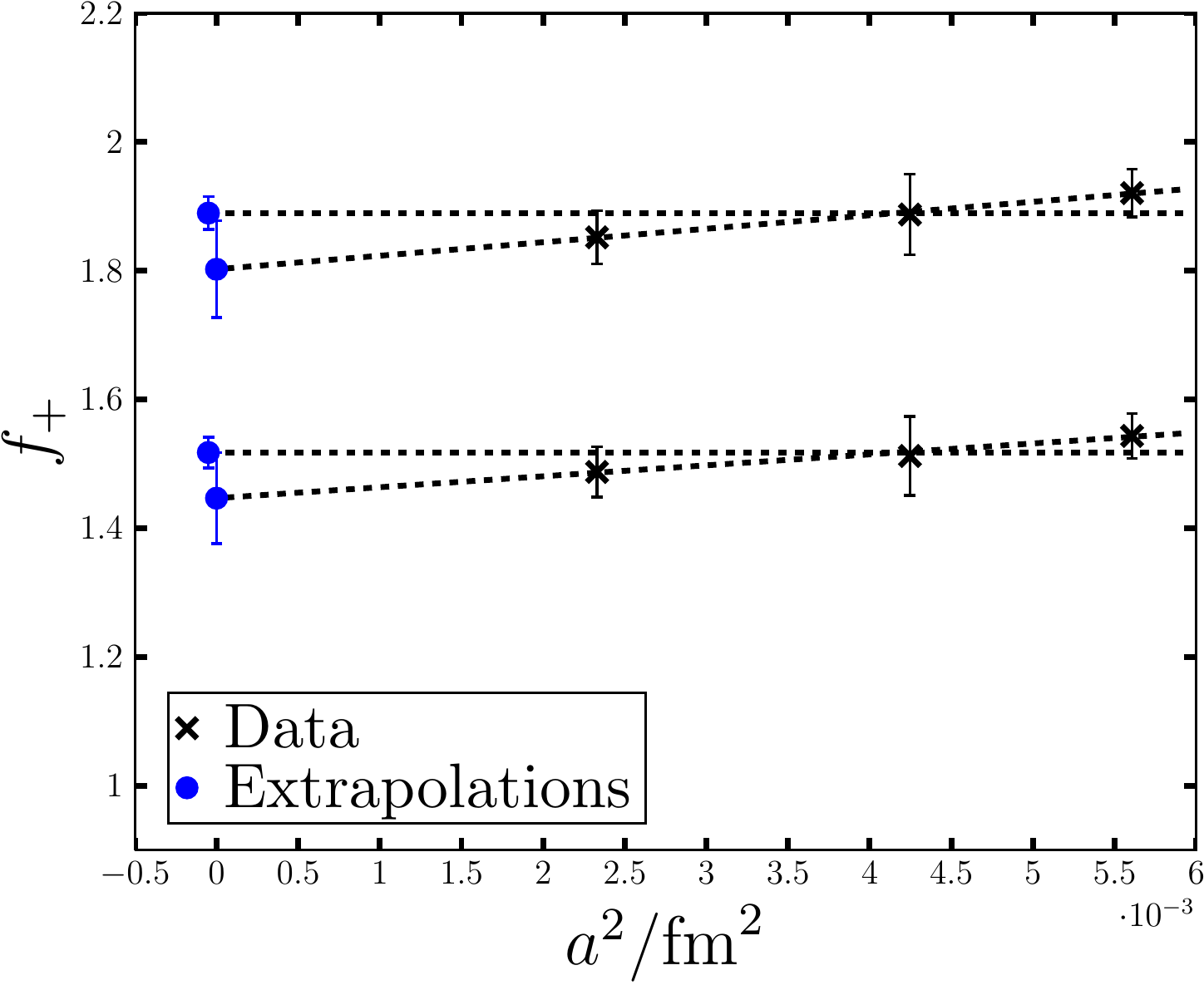}}
		\caption{Continuum extrapolation of our data
                  for $f_+$ at $q^2 = 21.23\,\mathrm{GeV}^2$.
		}
                \label{fig:cont}
\end{minipage}
\hfill
\begin{minipage}[hbt]{0.48\textwidth}
		\scalebox{0.5}{\includegraphics{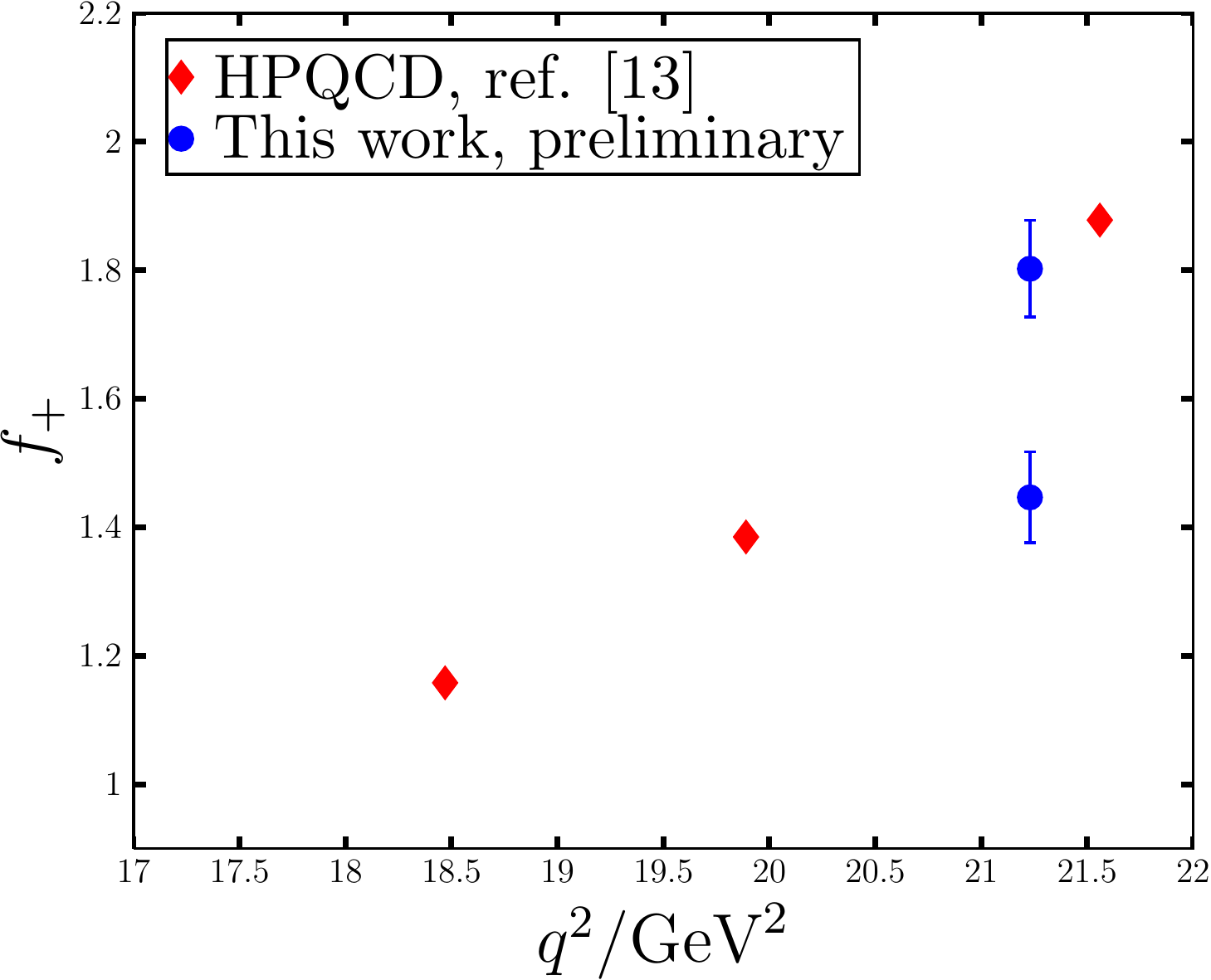}}
		\caption{Comparison of LQCD results at various values of $q^2$.}
		\label{fig:comp}
\end{minipage}
\end{figure}

Fig. \ref{fig:cont} shows $f_+$, as extracted from the fitted $\varphi^{(1,1)}_\mu$,
for different lattice spacings.
Working in the static approximation of HQET, we are free to keep or drop terms 
of order $1/m_\mathrm h$ in eq.~\eqref{fpl} for computing $f_+$. 
To illustrate this $O(1/m_\mathrm h)$ ambiguity, we show in fig.~\ref{fig:cont}
(and \ref{fig:comp}) two sets of data points:
the upper one corresponds to using all terms in eq.~\eqref{fpl},
the lower one to dropping the term proportional to $f_\parallel$.
Once we include all $O(1/m_\mathrm h)$ terms of HQET, this ambiguity
will disappear.
For both sets we show a constant continuum extrapolation and one linear in $a^2$.
The latter has by far the larger error and within this error is consistent with the 
result of the constant extrapolation.

In fig.~\ref{fig:comp}, we compare our results from the linear continuum extrapolation 
of $f_+(q^2)$ to recent results of HPQCD \cite{Bouchard:2014ypa} (at their smallest 
$a = 0.09\,\mathrm{fm}$ and $m_\pi = 320\, \mathrm{MeV}$).

\section{Conclusion}
%%%%%%%%%%%%%%%%%%%%
We presented the current status of our computation of the form factor 
$f_+(q^2)$ for the semi-leptonic decay $\mathrm B _\mathrm s \to \mathrm K \ell\nu$ 
at a fixed value of $q^2 = 21.23\,\mathrm{GeV}^2$ using HQET on the lattice. 
We compare two different methods to extract the form factors, either from 
the plateau value of a suitable ratio of correlators, or from a simultaneous
fit to the functional form of the correlators.

We also have performed a continuum extrapolation of our lattice data and find 
small $O(a^2)$ effects. The preliminary results reported here are still
computed in the static approximation and an extrapolation to the physical pion mass 
has yet to be performed. Our preliminary value of $f_+$ at this stage 
is in rough agreement with the results from other collaborations.

%%% The correlators to consistently include all $O(1/m_\mathrm h)$ HQET effects 
%%% have been measured and 
All $O(1/m_\mathrm h)$ effects of HQET
will be included in the analysis once the HQET parameters
are known non-perturbatively. We also plan to extend the computation to 
$\mathrm B\to\pi\ell\nu$ decays, several values of $q^2$, and $N_\mathrm f=2+1$
flavours of sea quarks.

%%%%%%%%%%%%%%%%%%%%%%%%%%%%%%%%%%%%%%%%%%%%%%%%%%%%%%%%%%%%%%%%%%%%%%%%%
\Acknowledgements
We thank the Leibniz Supercomputing Centre for providing computing time on SuperMUC.
The gauge configurations were produced by the CLS effort and for the used
computing resources we refer to the acknowledgement of ref.~\cite{Fritzsch:2012wq}.

\end{document}